\documentclass[
  nojss]{jss}

\usepackage{orcidlink,thumbpdf,lmodern}

\usepackage[utf8]{inputenc}

\author{
Xiaolei Wang~\orcidlink{0009-0005-6192-9061}\\University of
Melbourne \And Tomasz
Woźniak~\orcidlink{0000-0003-2212-2378}\\University of Melbourne
}
\title{Bayesian Analyses of Structural Vector Autoregressions with Sign,
Zero, and Narrative Restrictions Using the~\proglang{R}~Package
\pkg{bsvarSIGNs} \linebreak (Version 2.0)}

\Plainauthor{Xiaolei Wang, Tomasz Woźniak}
\Plaintitle{Bayesian Analyses of Structural Vector Autoregressions with
Sign, Zero, and Narrative Restrictions Using the R Package bsvarSIGNs}
\Shorttitle{\pkg{bsvarSIGNs}: Bayesian SVARs with Sign, Zero, and
Narrative Restrictions}

\Abstract{
\noindent The \proglang{R} package \pkg{bsvarSIGNs} implements
state-of-the-art algorithms for the Bayesian analysis of Structural
Vector Autoregressions identified by sign, zero, and narrative
restrictions. It offers fast and efficient estimation thanks to the
deployment of frontier econometric and numerical techniques and
algorithms written in \proglang{C++}. The core model is based on a
flexible Vector Autoregression with estimated hyper-parameters of the
Minnesota prior and the dummy observation priors. The structural model
can be identified by sign, zero, and narrative restrictions, including a
novel solution making it possible to use the three types of restrictions
at once. The package facilitates predictive and structural analyses
using impulse responses, forecast error variance and historical
decompositions, forecasting and conditional forecasting, as well as
analyses of structural shocks and fitted values. All this is
complemented by colourful plots, user-friendly summary functions, and
comprehensive documentation. The package was granted the Di~Cook
Open-Source Statistical Software Award by the Statistical Society of
Australia in 2024.
}

\Plainkeywords{Bayesian inference, Structural VARs, Gibbs sampler, sign
restrictions, zero restrictions, narrative
restrictions, forecasting, structural analysis, R}

\Address{
    Xiaolei Wang\\
    University of Melbourne\\
    Department of Economics\\
111 Barry Street\\
3053 Carlton, VIC, Australia\\
  E-mail: \email{adam.wang@unimelb.edu.au}\\
  URL: \url{https://github.com/adamwang15}\\~\\
      Tomasz Woźniak\\
    University of Melbourne\\
    Department of Economics\\
111 Barry Street\\
3053 Carlton, VIC, Australia\\
  E-mail: \email{tomasz.wozniak@unimelb.edu.au}\\
  URL: \url{https://github.com/donotdespair}\\~\\
  }

\providecommand{\tightlist}{%
  \setlength{\itemsep}{0pt}\setlength{\parskip}{0pt}}

\usepackage[utf8]{inputenc} \usepackage{tikz}

\usetikzlibrary{shapes,positioning,arrows.meta}

\usepackage{float} \usepackage[none]{hyphenat} \usepackage{amsmath,amsfonts,amssymb,amsthm,pxfonts,dsfont} \usepackage{natbib} \usepackage{graphicx}  \usepackage{color} \usepackage{geometry} \usepackage{multirow} \usepackage{booktabs} \usepackage{rotating}  \usepackage{longtable} \usepackage{subfig} \usepackage{wrapfig} \usepackage{subfig}

\emergencystretch=\maxdimen \hyphenpenalty=10000 \hbadness=10000

\begin{document}

\section{Introduction}\label{sec:intro}

Following the seminal developments proposed by \cite{Sims1980} and
\cite{Doan1984} Structural Vector Autoregressions (SVARs) have become
go-to models for empirical research in macroeconomics and finance. This
fact is attributed to the model's capacity to accommodate basic features
of macroeconomic aggregates, relative simplicity of implementation, and
interpretability facilitating a wide range of applications
\citep[see, for instance,][]{KL2017}. The SVARs are models that treat
all the variables and endogenously determined in a structural,
stochastic, dynamic system capturing the data serial autocorrelation, as
well as their temporal and contemporaneous interdependencies. Additional
assumptions of linearity and conditional Gaussianity lead to analytical
tractability and parsimonious estimation
\citep[see, for instance,][]{L2005,karlsson2013}. Finally, they can be
used to investigate Granger causality, dynamic causal effects of
well-isolated shocks, network connections, and forecast future values of
the variables of interest. All these features, make them indispensable
for decision-making at economic governance and financial institutions.

Bayesian approach applied to SVARs facilitates inference in finite
samples, which is particularly important due to a relatively small
number of observations given the large dimension of the parameter space
in many applications. In this context, the prior distribution of
parameters of the model that must be specified to apply Bayes' rule
provides an additional tool to handle identification, sparsity,
regularisation, and improve the forecasting performance. Additionally,
the reliability of numerical methods used for Bayesian estimation has
driven the incorporation of a large number of variables, sophisticated
identification strategies, and non-linearity in the model specification.
All this comes at a cost of computational complexity to be paid for
estimation, inference, and forecasting. Following these developments,
macroeconometrics as an academic field became increasingly computational
and characterised by data-driven application-specific modelling
approaches.

An important development facilitating Bayesian analysis of Vector
Autoregressions (VARs) was the application of a matrix-variate normal
inverse Wishart prior distribution for the model parameters, namely the
autoregressive parameters matrix and error term covariance matrix,
proposed by \cite{DM1976} for the multivariate regression model and
adapted to VARs by \cite{KK1997}. This natural-conjugate prior combined
with the assumption of conditional normality of the error term result in
a joint posterior distribution of these parameters being matrix-variate
normal inverse Wishart \citep[see, e.g.,][]{wozniak2016}. This posterior
distribution has a known analytical form, is easy to sample from, and
has become the basis for multiple modelling approaches
\citep[see, for instance,][]{Banbura2010,Koop2013,Chan2020}.

This paper and the corresponding \proglang{R} package \pkg{bsvarSIGNs}
by \cite{bsvarSIGNs} provides fast and efficient algorithms for Bayesian
analysis of SVARs. It takes advantage of this convenient model
formulation and applies frontier econometric and numerical techniques
and code written in \proglang{C++} to ensure blazingly fast
computations. Additionally, it offers a great flexibility in choosing
the model identification pattern, modifying the prior assumptions, and
accessing interpretable tabulated or plotted outputs. Therefore, the
package makes it possible to benefit from the best of the two
facilities: the convenience of data analysis using \proglang{R} and the
computational speed using code written in \proglang{C++}.

More specifically the \pkg{bsvarSIGNs} package builds upon a Bayesian
VAR model relying on the conditional prior and posterior distribution
for the parameter of the model being the normal-inverse Wishart
distribution. The models can be further extended by hierarchical prior
distribution as in \cite{GLP2015} featuring the Minnesota and
dummy-observation priors by \cite{Doan1984}, explicit specification of a
prior distribution for the rotation of the structural system as in
\cite{RRWZ2010}. These additional model features are introduced through
their corresponding marginal prior distributions and they are treated as
given in the conditional prior specification of the VAR parameters. This
construction of the joint prior and posterior distributions allow the
models to benefit from both: the analytical convenience of the natural
conjugate prior distribution granting computational efficiency and the
flexibility of the structural model specification leading to its
empirical relevance.

This convenient model construction facilitates the application of a
variety of methods for the identification of structural shocks. The
\pkg{bsvarSIGNs} package implements the identification of the SVARs
using sign restrictions as in \cite{RRWZ2010}, zero and sign
restrictions proposed by \cite{ARRW2018}, and narrative sign
restrictions introduced by \cite{ADRR2018}. We also prove that all of
these restrictions can be used simultaneously and include this novel
possibility in the package algorithms.

All of these models are set identified up to a rotation matrix such that
the sign, zero, and narrative restrictions hold. Appropriate handling of
the orthogonal matrix rotating the structural system is essential as
this rotation is not updated with the information from the data and is
featured in the asymptotic posterior distribution of the model
parameters as shown by \cite{BH2015}. Following \cite{RRWZ2010}, all the
models provided in the \pkg{bsvarSIGNs} package sample the rotation
matrix from the uniform distribution over the space of orthogonal
matrices, namely the Haar distribution by \cite{stewart1980}.
Subsequently, this draw, together with the VAR model parameters, is
accepted or not based on whether the sign, zero, or narrative
restrictions hold. When the zero and narrative restrictions are used,
the accepted draws are resampled using importance weights as in
\cite{ARRW2018} and \cite{ADRR2018}. This approach ensures that the
posterior distribution of the model parameters is consistent with the
restrictions imposed on the structural shocks and that the coverage of
the subspace of orthogonal matrices consistent with the restrictions is
optimal.

Additionally, the \pkg{bsvarSIGNs} package is aligned regarding objects,
workflows, and code structure with the \proglang{R} package \pkg{bsvars}
by \cite{bsvars,Wozniak2024} for the Bayesian SVARs identified by
exclusion restrictions, heteroskedasticity and non-normality, and they
constitute an integrated toolset with more information available at
\href{https://bsvars.org/}{https://bsvars.org}. Altogether, these
packages offer an unmatched number of SVAR models that will be
appropriate for the estimation of the effects of monetary and fiscal
policies, analysis of labour, financial, and housing markets, and many
other essential applications.

\begin{figure}
\centering
\includegraphics[width=2in,height=2.314in]{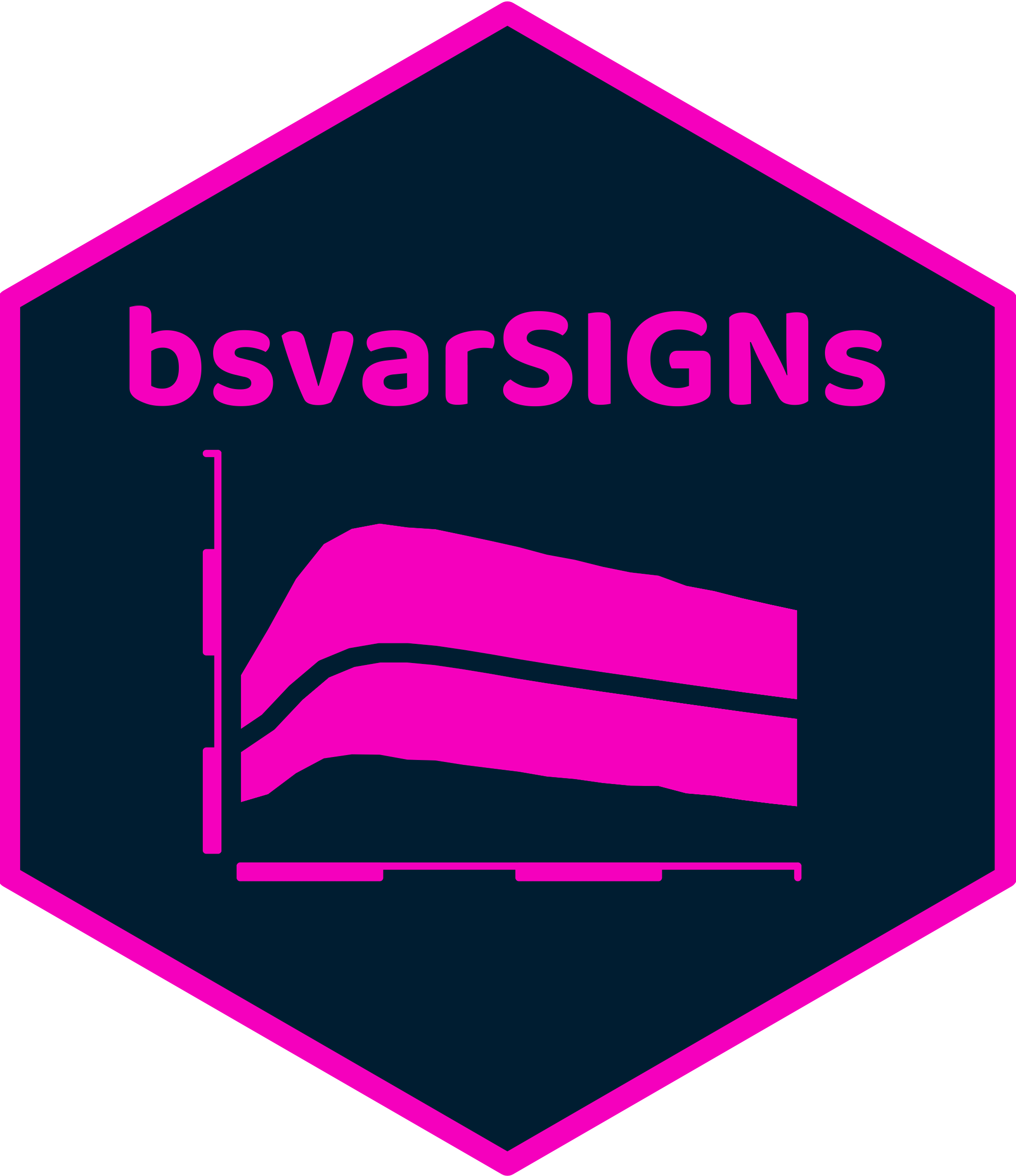}
\caption{The hexagonal package logo features an impulse response that
can be fully reproduced using the \pkg{bsvarSIGNs} package following a
script available at:
\href{https://github.com/bsvars/hex/blob/main/bsvarSIGNs/bsvarSIGNs.R}{github.com/bsvars/hex/blob/main/bsvarSIGNs/bsvarSIGNs.R}}
\end{figure}

The package implements also a wide range of tools for structural and
predictive analyses. The former encompasses the methods comprehensively
revised by \citep[][Chapter 4: SVAR Tools]{KL2017} and include the
impulse response functions, forecast error variance decompositions,
historical decompositions, as well as the basic analysis of fitted
values, and structural shocks. The predictive analysis includes Bayesian
forecasting implemented through an algorithm sampling from the
predictive density of the unknown future values of the dependent
variables, and conditional forecasting of a number of variables given
other variables' projections. Methods \texttt{summary()} and
\texttt{plot()} support the user in the interpretations and
visualisation of such analyses.

The \pkg{bsvarSIGNs} package is written for \proglang{R} \citep{Rcore}.
However, all of the algorithms are written in \proglang{C++} implemented
thanks to the \pkg{Rcpp} by \cite{eddelbuettel2011rcpp}, \cite{Rcpp},
and \cite{eddelbuettel_seamless_2013}. The essential parts of our
algorithms rely on linear algebra operations and random matrices
generators provided by the \proglang{C++} library \pkg{Armadillo} by
\cite{sanderson2016armadillo} through the \pkg{RcppArmadillo} package by
\cite{RcppArmadillo} and \cite{eddelbuettel_rcpparmadillo2014}. We use
the progress bar from package \pkg{RcppProgress} by \cite{RcppProgress}.
Finally, the \pkg{bsvarSIGNs} package provides a range of methods for
the generics defined by the \pkg{bsvars} package by \cite{bsvars} that
use classes from the \pkg{R6} package by \cite{R6}.

The \pkg{bsvarSIGNs} package offers a range of improvements relative to
existing software packages. Our algorithms are faster than those from
competing packages for the same models or procedures, and we offer a
greater range of model specifications and interpretable outputs to
report. As we have already pointed out this package is highly-compatible
with the \pkg{bsvars} package by \cite{bsvars} in terms of the
workflows, objects, and code structure. These two packages differ in
terms of the model structures leading to different classes of Bayesian
estimation algorithms. While the \pkg{bsvarSIGNs} package provides
models that lead to independent draws from the posterior distribution of
the parameters, the \pkg{bsvars} package relies on Monte Carlo Markov
Chain methods producing dependent samples.

The most similar existing software packages offering Bayesian SVARs are
the \proglang{R} package \pkg{BVAR} by \cite{BVAR} and the \pkg{BEAR}
toolbox by \cite{BEAR} for \proglang{MATLAB}. The former implements the
same hierarchical model and identification through sign and zero
restrictions. The latter implements the sign restrictions and a
selection of other structural models. None of them implements narrative
restrictions or the possibility of using sign, zero, and narrative
restrictions at once. \cite{Wozniak2024} revises other relevant packages
for VAR and SVAR models comprehensively. Here, we only mention those
relying on Bayesian inference such as \pkg{bvarsv} by \cite{bvarsv} and
\pkg{FAVAR} by \cite{FAVAR} implementing specific SVARs from influential
studies by \cite{Primiceri2005} and \cite{Bernanke2005} respectively,
and those applying hierarchical modelling to reduced-form VAR models,
such as \pkg{bvartools} by \cite{bvartools}, \pkg{bayesianVARs} by
\cite{bayesianVARs}, and \pkg{BGVAR} by \cite{BGVAR}.

In this context, the \pkg{bsvarSIGNs} package offers a unique
combination of features significantly enhancing the existing tools for
Bayesian SVAR analysis. This is achieved by the appropriate VAR model
specification, providing a number of structural identification
strategies aligned with that specification, application of dedicated
numerical methods, writing algorithms in \proglang{C++}, designing
user-friendly workflows, offering a range of methods for structural and
predictive empirical analyses and ways of tabular and visual
presentation of the results. All this makes the \pkg{bsvarSIGNs} package
an indispensable tool for empirical macroeconomic and financial
research.

\section{Workflows for SVAR analysis}\label{sec:workflow}

\begin{figure}[ht!]
\begin{center}
\begin{tikzpicture}

\node[draw,
        minimum width=4cm,
        minimum height=1cm] 
    (specify) 
    {\begin{tabular}{l}
    \textit{Specify a model}\\[1ex]
    \verb|specify_bsvarSIGN|\\[1ex] \small
    example:\\ \small
    \verb|data(optimism)|\\ \small
    \verb|spec = specify_bsvarSIGN$new(optimism)|
    \end{tabular}};

\node[draw,
    below=of specify,
        minimum width=4cm,
        minimum height=1cm] 
    (hyper) 
    {\begin{tabular}{l}
    \textit{Estimate prior hyper-parameters}\\[1ex]
    \verb|spec$prior$estimate_hyper(S = 1000)|
    \end{tabular}};

\node[draw,
      below=of hyper,
        minimum width=4cm,
        minimum height=1cm] 
    (estimate) 
    {\begin{tabular}{l}
    \textit{Estimate a model}\\[1ex]
    \verb|estimate|
    \end{tabular}};

\node[draw,
      below right=of estimate,
        minimum width=4cm,
        minimum height=1cm] 
    (forecast) 
    {\begin{tabular}{l}
    \textit{Forecast}\\[1ex]
    \verb|forecast|
    \end{tabular}};

\node[draw,
      below=of forecast,
        minimum width=4cm,
        minimum height=1cm] 
    (compute) 
    {\begin{tabular}{l}
    \textit{Compute interpretable quantities}\\[1ex]
    \verb|compute_conditional_sd|\\
    \verb|compute_fitted_values|\\
    \verb|compute_historical_decompositions|\\
    \verb|compute_impulse_responses|\\
    \verb|compute_structural_shocks|\\
    \verb|compute_variance_decompositions|
    \end{tabular}};

\draw[-latex] (specify) edge (hyper);
\draw[-latex] (hyper) edge (estimate);
\draw[-latex] (estimate) |- (forecast);
\draw[-latex] (estimate) |- (compute);

\draw[arrows={-Stealth[scale=0.8,inset=0pt]}] (specify) to [bend left=90] (estimate);
\end{tikzpicture}
\end{center}
\caption{Workflow for SVAR analysis}
\label{fig:workflow}
\end{figure}
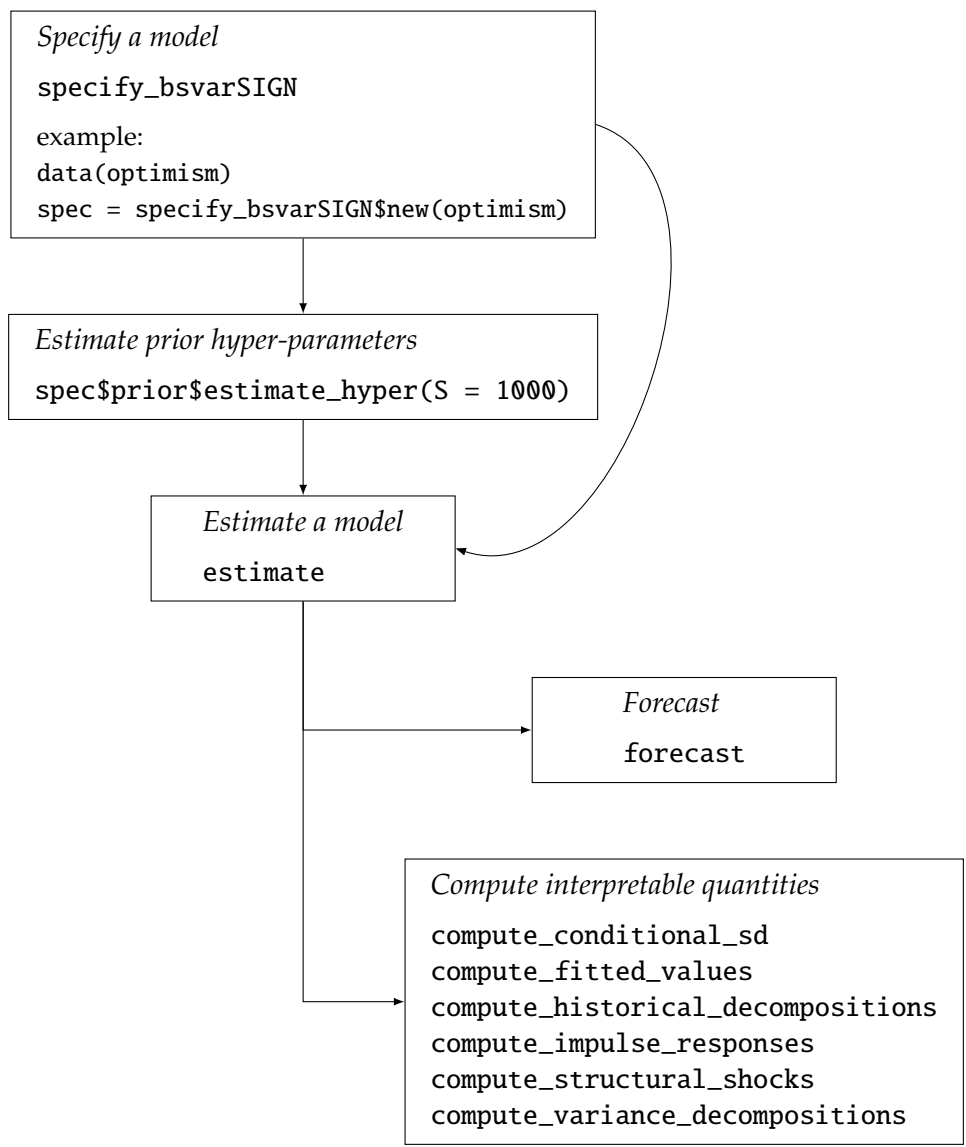

The methods proposed in this paper are implemented in the \proglang{R}
package \pkg{bsvarSIGNs} version 2.0 by \cite{bsvarSIGNs}. The package
offers a simple workflow for the model specification, estimation,
forecasting, computation of interpretable quantities, and their
summaries and visualizations. This section first presents the basic
workflow of the package visualized in Figure \ref{fig:workflow} and then
focuses on its particular steps.

Figure \ref{fig:workflow} illustrates the basic workflow of the
\pkg{bsvarSIGNs} package that is determined by the statistical
procedures. Working with the package begins with uploading the package
which uploads the sample data sets as well, including the matrix
\texttt{optimism} used by \cite{ARRW2018}. This data matrix is provided
as argument to the function \texttt{specify\_bsvarSIGN} that sets the
objects necessary to work with the model, such as the data matrices for
dependent and explanatory variables, prior hyper-parameters that are
fixed, identification restrictions, the VAR model lag order set to the
default value \texttt{p\ =\ 1}, starting values. They are set as
convenient \pkg{R6} objects.

The estimation proceeds in two steps. In the first optional one, the
hyper-parameters can be estimated using function
\texttt{estimate\_hyper} being a method of the \pkg{R6} objects
\texttt{spec} generated in the model specification by running
\texttt{specify\_bsvarSIGN}. The hyper-parameter estimation is done
marginally on the autoregressive parameters and the covariance matrix
following the model structure proposed by \cite{GLP2015}. In this step,
we use the Adaptive Random Walk Metropolis-Hastings algorithm by
\cite{am2006} and \cite{amh2010} as presented by
\cite{kalli2018bayesian}. It offers scalability, fast convergence, and
higher sampling efficiency than the original solution used by
\cite{GLP2015}. If the user decides not to estimate these
hyper-parameters and skip this step, the default values of
hyper-parameters proposed by \cite{GLP2015} are used.

Subsequently, for each of the hyper-parameter draws, or given the fixed,
that is, not estimated hyper-parameter values, the corresponding draws
of the autoregressive parameters and the covariance matrix are sampled
independently from the matrix-variate normal inverse Wishart
distribution \citep[see][]{wozniak2016} using method
\texttt{estimate.BSVARSIGN}. Here the autoregressive parameters are
drawn equation-by-equation as in \cite{CARRIERO2022} significantly
speeding up the computations.

Finally, given the draws of the parameters of the model, the user can
compute the corresponding posterior draws from the predictive density of
the future unknown values using method
\texttt{forecast.PosteriorBSVARSIGN}, or those for a selection of
interpretable quantities such as impulse responses, historical
decompositions, forecast error variance decompositions, as well as
conditional standard deviations, fitted values, and structural shocks
using the methods listed on the bottom of Figure \ref{fig:workflow}.
Each of these computations can be summarised in terms of the
corresponding posterior distribution mean, standard deviation, and
quantiles, and visualized using the methods \texttt{summary} and
\texttt{plot} respectively.

\subsection{The Basic Workflow}\label{sec:basic}

Upload the package to the \proglang{R} environment by executing the
following line:

\begin{CodeChunk}
\begin{CodeInput}
R> library(bsvarSIGNs)
\end{CodeInput}
\end{CodeChunk}

The package includes sample data object formatted so that they can be
provided directly as arguments of functions. The \code{optimism} data
set, downloaded from the replication package of \cite{ARRW2018},
contains the time series formatted as \texttt{matrix}. Upload this data
object to the memory and display several first observations by

\begin{CodeChunk}
\begin{CodeInput}
R> data(optimism)
R> head(optimism)
\end{CodeInput}
\begin{CodeOutput}
     productivity stock_prices consumption real_interest_rate hours_worked
[1,]    0.2172072    -11.28895   -4.331866        0.008022252    -7.599184
[2,]    0.2129482    -11.17647   -4.324255        0.021877748    -7.592445
[3,]    0.2069894    -11.11805   -4.318455        0.018811208    -7.580577
[4,]    0.2003908    -11.08317   -4.301382        0.012867114    -7.572133
[5,]    0.1945243    -11.02211   -4.293948        0.024503568    -7.577512
[6,]    0.2002138    -11.06306   -4.291474        0.000876887    -7.577491
\end{CodeOutput}
\end{CodeChunk}

To impose sign restrictions on the impulse response functions, the user
needs to specify a \(N\times N\times h\) sign restrictions array where
\(h\) is number of periods. When only imposing restrictions on the
contemporaneous responses, a \(N\times N\) matrix is also accepted. This
sign restriction array only accepts -1, 0, 1 and NA values. Specify -1
if the user wants to impose a negative sign restriction, 0 for a zero
restriction, 1 for a positive sign restriction, and NA if the user does
not want to impose any restrictions. In each matrix of the restriction
array, one can interpret rows as variables and columns as structural
shocks. For example, to identify the optimism shock as in
\cite{ARRW2018}, the user can impose restrictions on contemporaneous
impulse responses of the first shock by:

\begin{CodeChunk}
\begin{CodeInput}
R> sign_irf = matrix(c(0, 1, rep(NA, 23)), 5, 5)
R> sign_irf
\end{CodeInput}
\begin{CodeOutput}
     [,1] [,2] [,3] [,4] [,5]
[1,]    0   NA   NA   NA   NA
[2,]    1   NA   NA   NA   NA
[3,]   NA   NA   NA   NA   NA
[4,]   NA   NA   NA   NA   NA
[5,]   NA   NA   NA   NA   NA
\end{CodeOutput}
\end{CodeChunk}

Specify the model by running a simple function that specifies the model
and all the required values, such as the number of lags, data matrices,
prior distribution parameters, and some characteristics for the specific
steps of the estimation algorithm. For instance, the code below will use
data matrix \texttt{optimism} (by setting the first argument
\texttt{data\ =\ optimism}), specify a model with 4 lags
(\texttt{p\ =\ 4}), and the Minnesota prior with prior mean for the
autoregressive parameters reflecting unit-root non-stationary of the
variables (the default value of the argument \texttt{stationary} that is
not explicitly stated below), sign restrictions on the contemporaneous
impulse responses (\texttt{sign\_irf\ =\ sign\_irf}), and save the model
specification in the object \code{spec} that is of class
\code{BSVARSIGN}.

\begin{CodeChunk}
\begin{CodeInput}
R> spec = specify_bsvarSIGN$new(optimism, p = 4, sign_irf = sign_irf)
\end{CodeInput}
\end{CodeChunk}

This object contains all the necessary information for the estimation of
the model. In particular, following \cite{GLP2015} it includes
hyper-parameters \(\mu\) for the sum-of-coefficients prior, \(\delta\)
for the single-unit-root prior, as well as \(\lambda\) and \(\psi\) for
the Minnesota prior. By default, they are fixed to the values used by
\cite{Doan1984}.

In a fully Bayesian way, the user can also specify hyper-prior
distributions for these hyper-parameters, and sample from their
posterior distributions with an adaptive Metropolis algorithm using
\code{spec\$prior\$estimate_hyper()}. To sample 5,000 burn-in draws and
10,000 posterior draws in the returned sample for all hyper-parameters,
the user can run the following code:

\begin{CodeChunk}
\begin{CodeInput}
R> set.seed(123)
R> spec$prior$estimate_hyper(
+   S = 15000,
+   burn_in = 5000,
+   mu = TRUE,
+   delta = TRUE,
+   lambda = TRUE,
+   psi = TRUE
+ )
\end{CodeInput}
\end{CodeChunk}

Given hyper-parameters, the VAR model exploits the normal inverse
Wishart conjugate prior distribution, which allows for independent
sampling from the posterior distribution of the same form without
relying on MCMC methods. This is done by passing the \code{spec} object
as the first argument to the function \code{estimate()}, the following
code samples 10,000 independent draws from the posterior distribution
and displays the progress bar:

\begin{CodeChunk}
\begin{CodeInput}
R> set.seed(123)
R> post = estimate(spec, S = 10000)
\end{CodeInput}
\begin{CodeOutput}
**************************************************|
 bsvarSIGNs: Bayesian Structural VAR with sign,   |
             zero and narrative restrictions      |
**************************************************|
 Progress of simulation for 10000 independent draws
 Press Esc to interrupt the computations
**************************************************|
\end{CodeOutput}
\end{CodeChunk}

One way to interpret the structural parameters is to compute the impulse
response functions of the structural shocks. The user can compute the
impulse response functions and plot them by running the following code:

\begin{CodeChunk}
\begin{CodeInput}
R> irf = compute_impulse_responses(post, horizon = 20)
R> plot(irf, probability = 0.68, col = "#F500BD")
\end{CodeInput}


\begin{center}\includegraphics{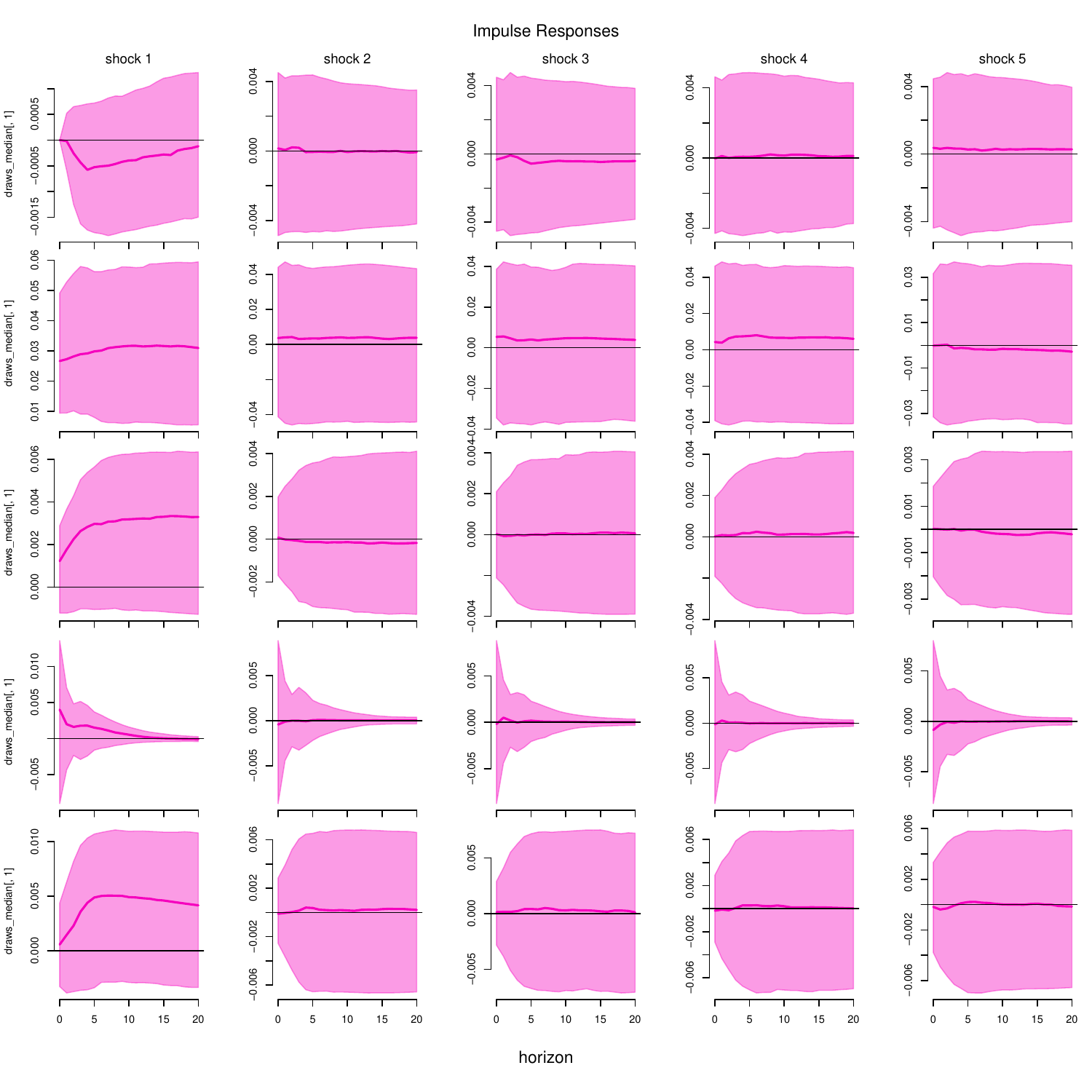} \end{center}

\end{CodeChunk}

This IRF plots replicates the results of \cite{ARRW2018} and shows the
response of the variables to a one standard deviation optimism shock. On
impact, we see the optimism shock does not affect productivity and there
is significant positive response of stock prices by construction, which
can be seen in the plot from the fifth row and the first column. The
response of the other variables, consumption, real interest rate, and
hours worked, are not significantly different from 0.

On the other hand, the user can also compute the forecasts from the
reduced-form parameters and plot them by running the following code:

\begin{CodeChunk}
\begin{CodeInput}
R> fore = forecast(post, horizon = 8)
R> plot(fore, data_in_plot = 0.2, col = "#F500BD")
\end{CodeInput}


\begin{center}\includegraphics{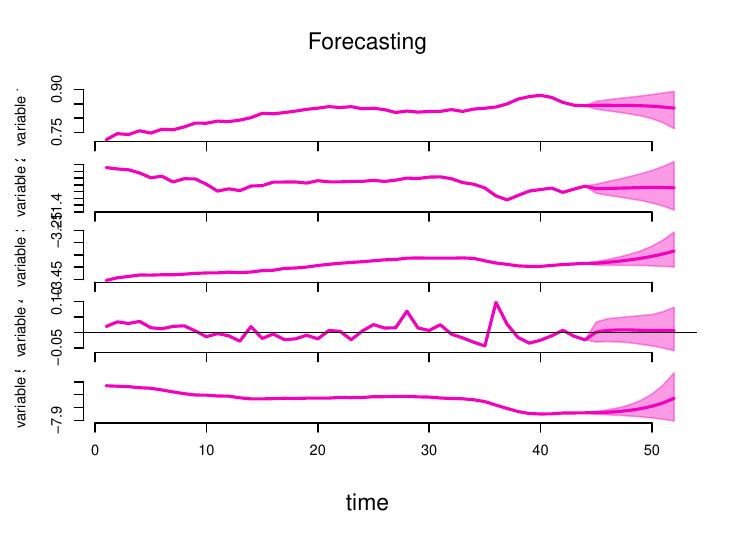} \end{center}

\end{CodeChunk}

In what follows, we focus on particular elements of the workflow
describe above, and other interpretations of the model result. Including
details on specifying the model, estimating hyper-parameters, estimating
the model, structural analysis and predictive analysis.

\subsection{Install the Package}

The package is available on CRAN and can be installed using the
following command:

\begin{CodeChunk}
\begin{CodeInput}
R> install.packages("bsvarSIGNs")
\end{CodeInput}
\end{CodeChunk}

To install the development version from GitHub repository
\href{https://github.com/bsvars/bsvarSIGNs}{github.com/bsvars/bsvarSIGNs},
use the following command:

\begin{CodeChunk}
\begin{CodeInput}
R> devtools::install_github("bsvars/bsvarSIGNs")
\end{CodeInput}
\end{CodeChunk}

The correct functioning of the \pkg{bsvarSIGNs} package requires the
installation of the \proglang{R} package \pkg{bsvars} by \cite{bsvars},
which will proceed automatically during the installation of the
\pkg{bsvarSIGNs} package.

Before every use of the package load it to the memory by running the
command:

\begin{CodeChunk}
\begin{CodeInput}
R> library(bsvarSIGNs)
\end{CodeInput}
\end{CodeChunk}

\subsection{Upload Data}

This package includes the \code{optimism} data set, downloaded from the
replication package of \cite{ARRW2018}. It contains the quarterly
observations from 1955 Q1 to 2004 Q4 of five variables in the United
States:

\begin{enumerate}
\def\labelenumi{\arabic{enumi}.}
\tightlist
\item
  Productivity: quarterly factor-utilization-adjusted total factor
  productivity.
\item
  Stock prices: quarterly end-of-period S\&P 500 divided by CPI.
\item
  Consumption: quarterly real consumption expenditures on nondurable
  goods and services.
\item
  Real interest rate: quarterly real interest rate.
\item
  Hours worked: quarterly hours of all persons in the non-farm business
  sector.
\end{enumerate}

To use the time series, load and visualise them by running the code:

\begin{CodeChunk}
\begin{CodeInput}
R> data(optimism)
R> plot(optimism, col = "#F500BD", nc = 1, bty = "n", cex.lab = 0.45)
\end{CodeInput}


\begin{center}\includegraphics{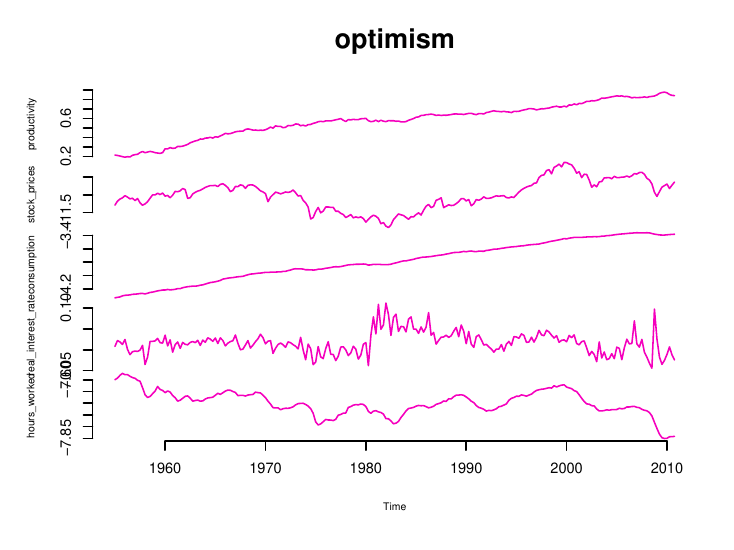} \end{center}

\end{CodeChunk}

Finally, more info about the data and object is available in the
documentation accessible by running \texttt{?optimism}.

\subsection{Specify the Model}

In specifying the model, there are two main steps. The first step is to
specify the sign restrictions. This package allows for three types of
restrictions: (i) restrictions on the impulse response functions, (ii)
narrative restrictions, and (iii) restrictions on the structural matrix
\(B\).

To specify (i) restrictions on the impulse response functions, the user
needs to specify the sign restrictions array as described in Section
\ref{sec:basic}. For example, to identify the optimism shock as in
\cite{ARRW2018}, the user can impose restrictions on contemporaneous
impulse responses of the first shock by:

\begin{CodeChunk}
\begin{CodeInput}
R> sign_irf = matrix(c(0, 1, rep(NA, 23)), 5, 5)
R> sign_irf
\end{CodeInput}
\begin{CodeOutput}
     [,1] [,2] [,3] [,4] [,5]
[1,]    0   NA   NA   NA   NA
[2,]    1   NA   NA   NA   NA
[3,]   NA   NA   NA   NA   NA
[4,]   NA   NA   NA   NA   NA
[5,]   NA   NA   NA   NA   NA
\end{CodeOutput}
\end{CodeChunk}

To specify (ii) narrative restrictions, the user needs to create a list
of \code{narrative} objects using the function
\code{specify_narrative()}. As in \cite{ADRR2018}, we allow for
narrative restrictions on the structural shocks and the historical
decompositions. For example, to impose a negative optimism shock during
the financial crisis in 2008 Q3, the user can specify the narrative
restriction by:

\begin{CodeChunk}
\begin{CodeInput}
R> date_fc        = which(time(optimism) == 2008.5)
R> narrative1     = specify_narrative(start = date_fc, sign = -1)
R> sign_narrative = list(narrative1)
\end{CodeInput}
\end{CodeChunk}

To specify (iii) restrictions on the structural matrix, the user needs
to create a \(N\times N\) matrix. Similar to (i), this matrix only
accepts values \texttt{-1}, \texttt{1} and \texttt{NA}. Use \texttt{-1}
to impose a negative sign restriction, \texttt{1} for a positive sign
restriction, and \texttt{NA} if the user does not want to impose any
restrictions for the particular period.

The second step in specifying the model is to create the specification
object using \code{specify_bsvarSIGN$new()}, which takes the data
object, the number of lags of the VAR model, the specified sign
restrictions, and other additional information. For example, to specify
a VAR model with 4 lags, sign restrictions on the impulse response
functions and narrative restrictions for structural shocks, the user can
run the following code:

\begin{CodeChunk}
\begin{CodeInput}
R> spec = specify_bsvarSIGN$new(
+   data           = optimism, 
+   p              = 4,
+   sign_irf       = sign_irf, 
+   sign_narrative = sign_narrative
+ )
\end{CodeInput}
\end{CodeChunk}

Finally, the documentation for the model specification can be accessed
by running \code{?specify_bsvarSIGN} with details on its particular
parts, such as the function generating the model specification
\code{specify_bsvarSIGN$new()}, available following the links in the
documentation just displayed.

\subsection{Estimate Hyper-Parameters}

This specification object \code{spec} contains all the necessary
information for the estimation of the model. In particular, following
\cite{GLP2015} it includes hyper-parameters \(\mu\) for the
sum-of-coefficients prior, \(\delta\) for the single-unit-root prior,
and \(\lambda\) and \(\psi\) for the Minnesota prior. By default, they
are fixed as \(\mu = 1,\ \delta = 1,\ \lambda = 0.2\), and \(\psi\) are
fixed as the OLS error variance estimates \citep[see][]{Doan1984}.

Before sampling the hyper-parameters, the user can display and alter the
prior distribution of the hyper-parameters by changing the default
values. The hyper prior of \(\psi\) follows an inverse gamma
distribution and the hyper prior of other hyper-parameters follows a
gamma distribution. For example, to display the prior distribution of
the shrinkage parameter \(\lambda\) in the Minnesota prior, the user can
run the following code:

\begin{CodeChunk}
\begin{CodeInput}
R> spec$prior$lambda.shape
\end{CodeInput}
\begin{CodeOutput}
[1] 1.370156
\end{CodeOutput}
\begin{CodeInput}
R> spec$prior$lambda.scale
\end{CodeInput}
\begin{CodeOutput}
[1] 0.5403124
\end{CodeOutput}
\end{CodeChunk}

These values can be changed by simply assigning a new value to them. For
example, \code{spec\$prior\$lambda.scale = 0.5} will change the scale
parameter of the prior distribution of \(\lambda\) accordingly.

Given the specification object \code{spec}, we are now ready to estimate
the hyper-parameters. Calling the function
\code{spec\$prior\$estimate_hyper()} will sample from the posterior
distribution of the hyper-parameters with an adaptive Metropolis
algorithm. The user can specify the total number of draws from the
posterior distribution by setting the argument \code{S}, and the number
of burn-in draws by setting the argument \code{burn_in} to be discarted,
then the algorithm will return \code{S - burn_in} draws of the
hyper-parameters. It is also possible to only estimate a subset of the
hyper-parameters by changing the argument of specific hyper-parameters
to \code{TRUE} or \code{FALSE}. For example, to sample 15,000 draws for
all the hyper-parameters and discard the first 5,000 draws as burn-in,
the user can run the following code:

\begin{CodeChunk}
\begin{CodeInput}
R> set.seed(123)
R> spec$prior$estimate_hyper(
+   S = 15000, 
+   burn_in = 5000,
+   mu = TRUE, 
+   delta = TRUE, 
+   lambda = TRUE, 
+   psi = TRUE
+ )
\end{CodeInput}
\begin{CodeOutput}
**************************************************|
 Adaptive Metropolis MCMC: hyper parameters       |
**************************************************|
\end{CodeOutput}
\begin{CodeInput}
R> hyper_draws = t(spec$prior$hyper)
R> colnames(hyper_draws) = c("mu", "delta", "lambda", paste0("psi", 1:ncol(optimism)))
R> plot.ts(hyper_draws, col = "#F500BD", bty = "n", xlab = "iterations")
\end{CodeInput}


\begin{center}\includegraphics{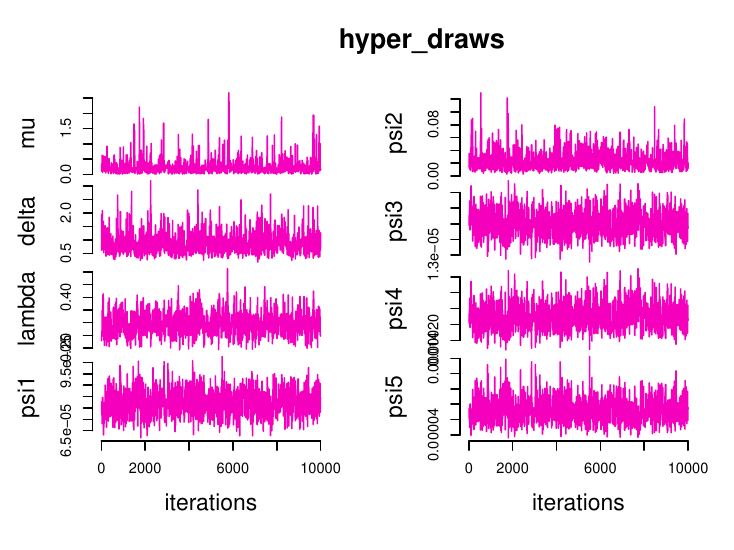} \end{center}

\end{CodeChunk}

The package also allows for more flexible specifications of the prior
distribution, which can be achieved by modifying the elements of the
\code{spec\$prior} list. For example, if the user wishes to replicate
the results in \cite{ARRW2018}, the specification should indicate fixed
hyper-parameters and turn off the sum-of-coefficients and
single-unit-root dummy observation priors. To do so, the user should set
the corresponding dummy observation matrices to have zero columns prior
to running the previous code:

\begin{CodeChunk}
\begin{CodeInput}
R> spec$prior$Ysoc = matrix(NA, nrow(spec$prior$Ysoc), 0)
R> spec$prior$Xsoc = matrix(NA, nrow(spec$prior$Xsoc), 0)
R> spec$prior$Ysur = matrix(NA, nrow(spec$prior$Ysur), 0)
R> spec$prior$Xsur = matrix(NA, nrow(spec$prior$Xsur), 0)
\end{CodeInput}
\end{CodeChunk}

\subsection{Estimate the Model}

Given hyper-parameters, whether fixed or a sample from the posterior
distribution, the user can then sample the VAR parameters. The VAR
models considered in the package all falls in the class of
normal-inverse Wishart conjugate prior models, which allows for
independent sampling from the posterior distribution without relying on
Monte Carlo Markov Chain methods. This is done by passing the
\code{spec} object as the first argument to the function
\code{estimate()}, the following code samples 10,000 independent draws
from the posterior distribution:

\begin{CodeChunk}
\begin{CodeInput}
R> set.seed(123)
R> post = estimate(spec, S = 10000)
\end{CodeInput}
\begin{CodeOutput}
**************************************************|
 bsvarSIGNs: Bayesian Structural VAR with sign,   |
             zero and narrative restrictions      |
**************************************************|
 Progress of simulation for 10000 independent draws
 Press Esc to interrupt the computations
**************************************************|
\end{CodeOutput}
\end{CodeChunk}

To assess the efficiency of the sampling algorithm, the user can print
out the effective sample size of the draws by running the following
code:

\begin{CodeChunk}
\begin{CodeInput}
R> post$posterior$ess
\end{CodeInput}
\begin{CodeOutput}
[1] 257.3824
\end{CodeOutput}
\end{CodeChunk}

\subsection{Report Structural Analysis}

The method \code{compute_structural_shocks()} computes the structural
shocks by appropriate rotations of the reduced-form shocks based on the
specified restrictions. To visualize the structural shocks, the user can
run the following code:

\begin{CodeChunk}
\begin{CodeInput}
R> ss = compute_structural_shocks(post)
R> plot(ss, col = "#F500BD")
\end{CodeInput}


\begin{center}\includegraphics{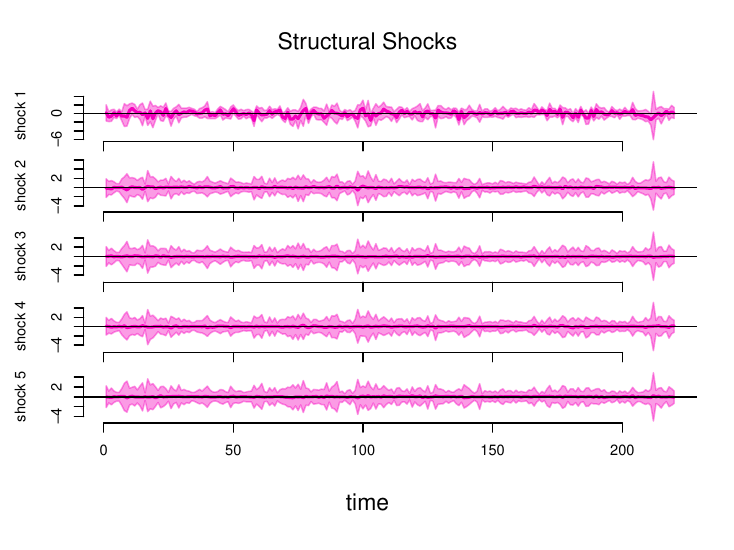} \end{center}

\end{CodeChunk}

To compute the impulse response functions, use the method
\code{compute_impulse_responses()} with the posterior object \code{post}
as the first argument, and specify the number of horizons
\code{horizon}. For example, to compute the impulse response functions
on impact and for 20 periods ahead, and visualize the result, run the
following code:

\begin{CodeChunk}
\begin{CodeInput}
R> irf = compute_impulse_responses(post, horizon = 20)
R> plot(irf, probability = 0.68, col = "#F500BD")
\end{CodeInput}


\begin{center}\includegraphics{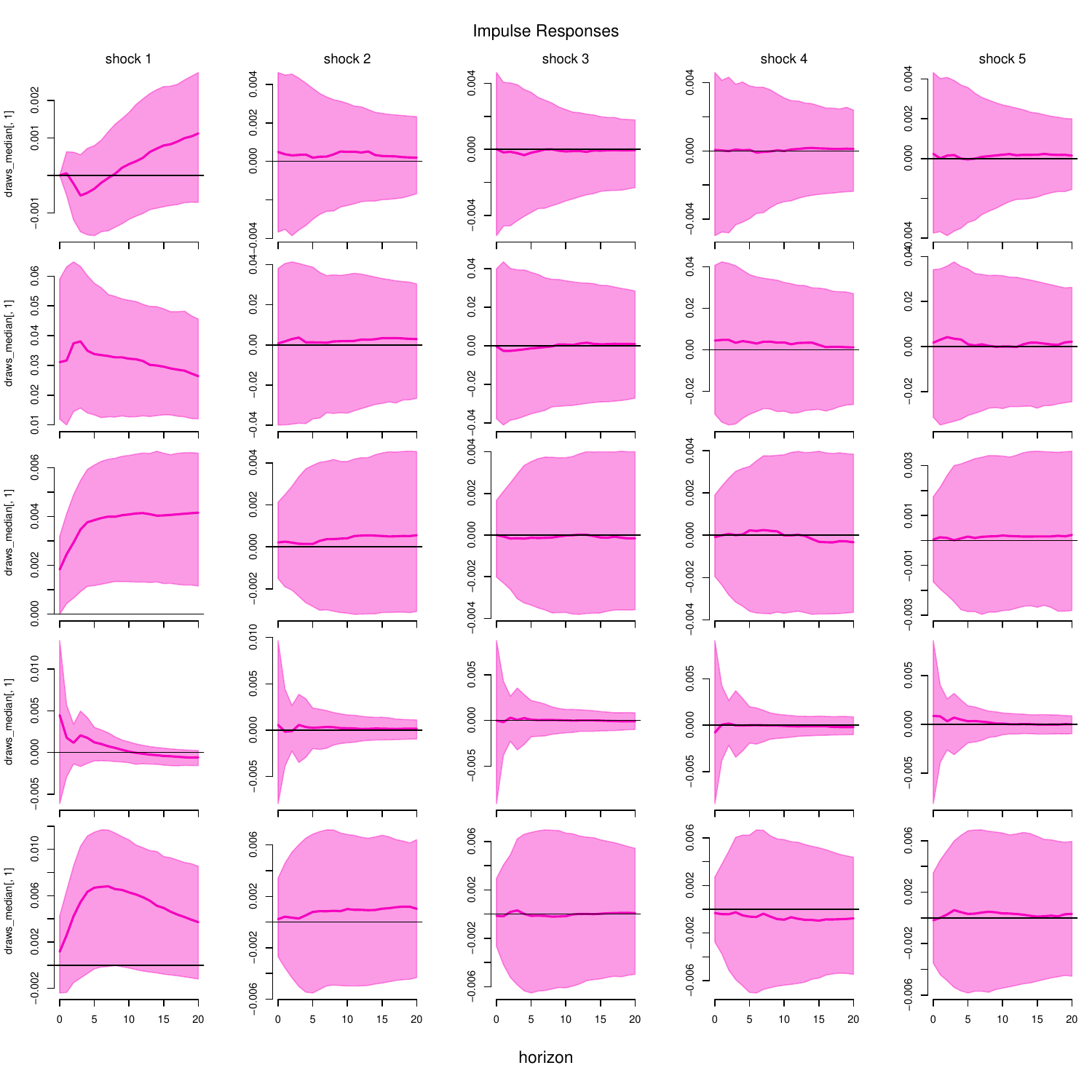} \end{center}

\end{CodeChunk}

This plot verifies the sign restrictions imposed on the contemporaneous
responses of the variables to the identified structural shocks. The
optimism shock (shock 1) has no contemporaneous effect on productivity
(variable 1) but has positive impact on stock prices (variable 2). The
effect of the optimism shock on both consumption and hours worked is
positive but not significant in the short run. However, a noticeable
sharpening of the identification of these effects is observed in a model
with the additional narrative restriction relative to the model with
only sign and zero restrictions reported in Section \ref{sec:basic}.

To compute the forecast error variance decompositions, use the method
\code{compute_variance_decompositions()} with the posterior object
\code{post} as the first argument. The user can then visualize the
forecast error variance decompositions by running the following code:

\begin{CodeChunk}
\begin{CodeInput}
R> fevd = compute_variance_decompositions(post, horizon = 20)
R> fc   = colorRampPalette(c("#F500BD", "#001D31"))
R> cols = fc(5)
R> plot(fevd, cols = cols)
\end{CodeInput}


\begin{center}\includegraphics{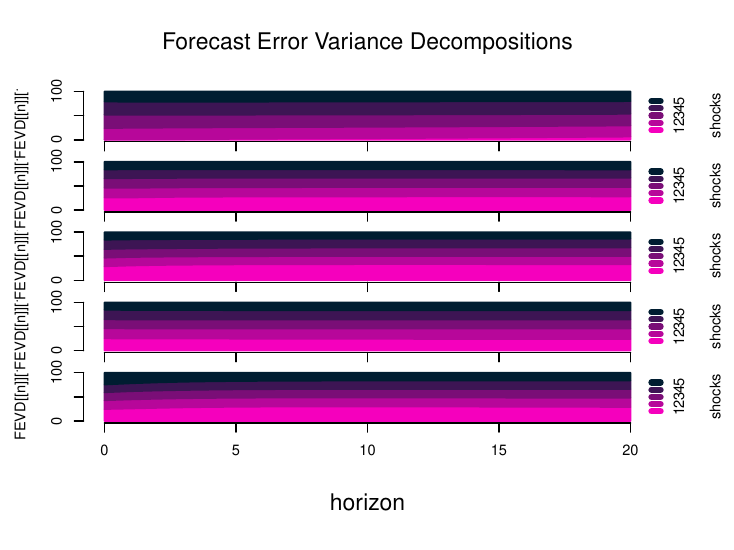} \end{center}

\end{CodeChunk}

\subsection{Report Predictive Analysis}

The method \code{compute_fitted_values()} computes the fitted values
from the reduced-form VAR model. More precisely, the method samples
random draws from the sample data density. TO perform th sampling and
visualise the in-sample model fit, the user can run the following code:

\begin{CodeChunk}
\begin{CodeInput}
R> fit = compute_fitted_values(post)
R> plot(fit, col = "#F500BD")
\end{CodeInput}


\begin{center}\includegraphics{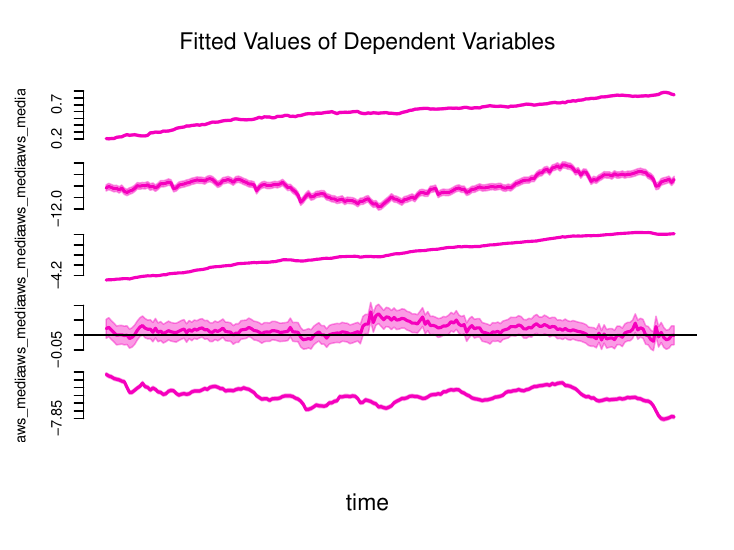} \end{center}

\end{CodeChunk}

To compute an 2-year out-of-sample forecast, the user can pass the
posterior object \code{post} as the first argument to the method
\code{forecast()} and specify the forecast horizon by setting, for
instancce, \code{horizon = 8}. In visualizing the forecast, it is
possible to focus on the recent periods by setting the argument
\code{data_in_plot} to be the desired percentage of the last
observations to be included in the plot. For example, to visualize the
forecast and the most recent 20\% of the data, the user can run the
following code:

\begin{CodeChunk}
\begin{CodeInput}
R> fore = forecast(post, horizon = 8)
R> plot(fore, data_in_plot = 0.2, col = "#F500BD")
\end{CodeInput}


\begin{center}\includegraphics{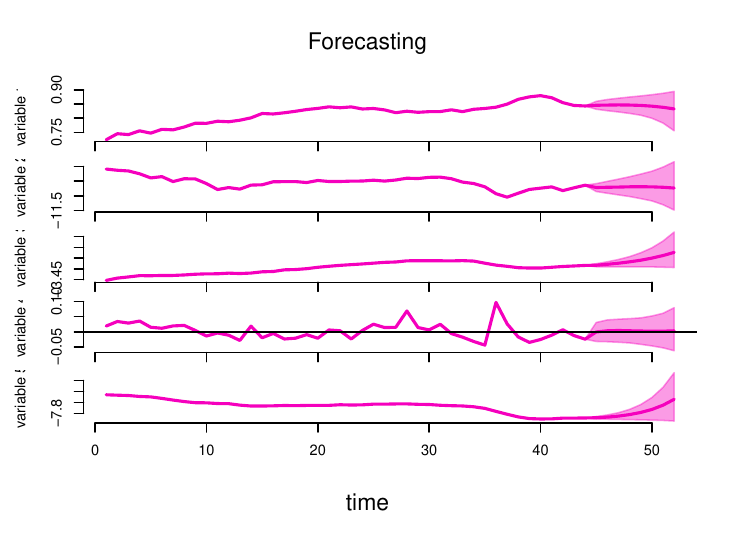} \end{center}

\end{CodeChunk}

The forecasts, and other interpretable outputs, can be also presented
using the \code{summary()} function, which provides a detailed table of
the forecasts. For example, to print out the forecast of productivity
(variable 1), the user can run the following code:

\begin{CodeChunk}
\begin{CodeInput}
R> summary(fore)$variable1
\end{CodeInput}
\begin{CodeOutput}
 **************************************************|
 bsvars: Bayesian Structural Vector Autoregressions|
 **************************************************|
   Posterior summary of forecasts                  |
 **************************************************|
\end{CodeOutput}
\begin{CodeOutput}
       mean          sd 5% quantile 95% quantile
1 0.8458038 0.008377681   0.8317997    0.8595328
2 0.8463867 0.011949591   0.8264638    0.8661054
3 0.8467017 0.014756666   0.8220349    0.8707619
4 0.8463071 0.017534978   0.8168991    0.8749356
5 0.8449208 0.020845072   0.8105439    0.8788074
6 0.8422181 0.025390293   0.7998666    0.8831610
7 0.8374115 0.032229660   0.7828341    0.8881583
8 0.8297283 0.042756979   0.7555310    0.8947330
\end{CodeOutput}
\end{CodeChunk}

\newpage
\section{Conclusions}\label{sec:conclusions}

The package \pkg{bsvarSIGNs} provides a comprehensive set of tools for
estimating and analysing Bayesian structural vector autoregressive
models with sign, zero, and narrative restrictions. The package is
designed to be user-friendly and flexible, allowing users to specify a
wide range of prior distributions and restrictions. It implements
efficient algorithms with \proglang{C++} for estimation and inference,
including an adaptive Metropolis algorithm for sampling the
hyper-parameters and independent sampling for the VAR parameters. The
package also provides a set of functions to summarise and visualise the
results, including structural shocks, impulse response functions,
historical decompositions, forecast error variance decompositions,
fitted values, and forecasts.

\bibliography{bib}

\end{document}